\documentclass[aps,prl,twocolumn,superscriptaddress]{revtex4-1}
\usepackage{graphicx}
\usepackage{subfigure}
\usepackage{amsmath}
\usepackage{multirow}
\usepackage{mathrsfs}
\usepackage{hyperref}
\usepackage{slashed}
\usepackage{multirow}

\begin{document}

\title{$K_L-K_S$ mass difference from lattice QCD}

\newcommand\bnl{Brookhaven National Laboratory, Upton, NY 11973, USA}
\newcommand\cu{Physics Department, Columbia University, New York,
      NY 10027, USA}
\newcommand\riken{RIKEN-BNL Research Center, Brookhaven National
      Laboratory, Upton, NY 11973, USA}
\newcommand\soton{School of Physics and Astronomy, University of
  Southampton,  Southampton SO17 1BJ, UK}

\author{Z.~Bai}\affiliation{\cu}
\author{N.H.~Christ}\affiliation{\cu}
\author{T.~Izubuchi}\affiliation{\bnl}\affiliation{\riken}
\author{C.T.~Sachrajda}\affiliation{\soton}
\author{A.~Soni}\affiliation{\bnl}
\author{J.~Yu}\affiliation{\cu} 
\collaboration{RBC and UKQCD}


\begin{abstract}
We report on the first complete calculation of the $K_L-K_S$ mass difference, $\Delta M_K$, using lattice QCD.  The calculation is performed on a 2+1 flavor, domain wall fermion ensemble with a 330~MeV pion mass and a 575~MeV kaon mass. We use a quenched charm quark with a 949~MeV mass to implement Glashow-Iliopoulos-Maiani cancellation.  For these heavier-than-physical particle masses, we obtain $\Delta M_K =3.19(41)(96)\times 10^{-12}$~MeV, quite similar to the experimental value.  Here the first error is statistical and the second is an estimate of the systematic discretization error.  An interesting aspect of this calculation is the importance of the disconnected diagrams, a dramatic failure of the OZI rule.
\end{abstract}


\maketitle

\section{Introduction}

The $K_L-K_S$ mass difference $\Delta M_K$, with a value of $3.483(6) \times 10^{-12}$~MeV~\cite{Nakamura:2010zzi}, is an important quantity in particle physics which led to the prediction of the energy scale of the charm quark nearly fifty years ago~\cite{Mohapatra:1968zz, Glashow:1970gm,Gaillard:1974hs} and whose small size places strong constraints on possible new physics beyond the standard model. This mass difference is believed to arise from $K^0$-$\overline{K}^0$ mixing caused by second-order weak interactions. However, because $\Delta M_K$ is suppressed by 14 orders of magnitude compared to the energy scale of the strong interactions and must involve a change in strangeness of two units, this is a promising quantity to reveal new phenomena which lie outside the standard model.  A quantity closely related to $\Delta M_K$ is the indirect $CP$ violation parameter $\epsilon_K$, which arises in the same mixing process. The experimental values of $\Delta M_K$ and $\epsilon_K$ are both known very accurately, making the precise calculation of $\Delta M_K$ and $\epsilon_K$ within the standard model an important challenge.  

As an example of new physics, consider a process which occurs with unit strength but at a very high energy scale $\Lambda$ and which changes strangeness by two units.  Such a process might be represented at low energies as the $\Delta S=2$, four-fermion operator $\frac{1}{\Lambda^2}\overline{s}d\overline{s}d$ where $\overline{s}$ and $d$ are operators creating a strange quark and destroying a down quark, respectively.  Establishing the validity of the standard model prediction for $\Delta M_K$ at the 10\% level would then provide a lower bound on $\Lambda$: $\Lambda \ge 10^4$~TeV -- an energy scale four orders of magnitude greater than is effectively available in present laboratory experiments.

In perturbation theory, the standard model contribution to $\Delta M_K$ is separated into short distance and long distance parts. The short distance part receives the largest contribution from momenta on the order of the charm quark mass. In the recent NNLO perturbation theory calculation of Brod and Gorbahn~\cite{Brod:2011ty}, the NNLO terms were found to be as large as 36\% of the leading order (LO) and next-to-leading order (NLO) terms, raising doubts about the convergence of the perturbation series at this energy scale.  At present the long distance part of $\Delta M_K$ is even less certain, with no available results with controlled errors because the long-distance contributions are non-perturbative.  However, an estimate given by Donoghue {\it et al.}~\cite{Donoghue:1983hi} suggests that the long distance contributions may be sizable. 

The calculation of $\epsilon_K$ is under much better control, because it is CP violating and the largest contribution involves momenta on the scale of top quark mass, where perturbation theory should be reliable. However, the same NNLO difficulties in predicting the charm quark contribution to $\epsilon_K$ enters at the 8\% level~\cite{Brod:2011ty}.  In addition the long distance contribution to $\epsilon_K$ is estimated to be 3.6\% by Buras {\it et al.}~\cite{Buras:2010pza}, again suggesting the need for a reliable, non-perturbative method.  Here long- and short-distances refer to the space-time separation between the two point-like, $\Delta S=1$ weak operators which enter the calculation of $\Delta M_K$ or $\epsilon_K$ when the internal loop momenta are much less than the $W$ boson mass.  Conventionally separations on the scale of $1/\Lambda_{\mathrm{QCD}}$ are referred to as ``long-distance''.

Lattice QCD provides a first-principles method to compute non-perturbative QCD effects in electroweak processes, in which all errors can be systematically controlled. We have proposed a lattice method to compute $\Delta M_K$ and $\epsilon_K$~\cite{Christ:2010zz,Christ:2012np}. An exploratory calculation of $\Delta M_K$ ~\cite{Christ:2012se} has been carried out on a 2+1 flavor, $16^3\times 32$, DWF ensemble with an unphysically large, 421~MeV pion mass. We obtained a mass difference $\Delta M_K$ which ranged from $6.58(30)\times 10^{-12}$~MeV to $11.89(81) \times 10^{-12}$~MeV for kaon masses varying from 563~MeV to 839~MeV. This exploratory work was incomplete since we included only a subset of the necessary diagrams.  

In this letter, we report on a full calculation, including all diagrams, with a lighter pion mass, larger volume and improved statistics.  The large lattice spacing and unphysical quark masses used in the calculation presented here prevent the resulting $\Delta M_K$  from being viewed as a test of the standard model.  However, this calculation demonstrates that a realistic lattice calculation of $\Delta M_K$ should be possible within a few years.  This calculation of amplitudes containing two effective weak operators represents an important advance in lattice technique and should allow future calculation of long distance effects in rare kaon decays and, possibly, heavy quark processes.

\section{Evaluation of $\Delta M_K$}
\begin{figure}[htp!]
	\includegraphics[width=0.4\textwidth]{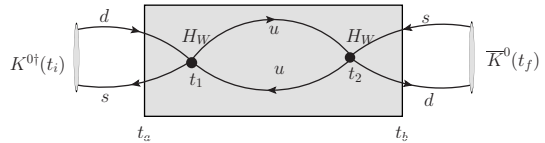}
\caption{One type of diagram contributing to $\mathscr{A}$ in Eq.~\eqref{eq:integrated_correlator}. Here $t_2$ and $t_1$ are integrated over the time interval $[t_a,t_b]$, represented by the shaded region.}
	\label{fig:int_correlator}
\end{figure}

We begin by summarizing the lattice method for evaluating $\Delta M_K$~\cite{Christ:2012se}. The essential step is to integrate the time-ordered product of two first-order weak Hamiltonians over a fixed space-time volume:
\begin{equation}
\mathscr{A}=\frac{1}{2}\sum_{t_2=t_a}^{t_b}\sum_{t_1=t_a}^{t_b}\langle0|T\left\{\overline{K}^0(t_f)H_W(t_2)H_W(t_1)\overline{K}^0(t_i)\right\}|0\rangle.
\label{eq:integrated_correlator}
\end{equation}
A class of diagrams contributing to this integrated correlator is represented schematically in Fig.~\ref{fig:int_correlator}.  After inserting a sum over intermediate energy eigenstates and summing explicitly over $t_2$ and $t_1$ in the interval $[t_a,t_b]$ one obtains:
\begin{eqnarray}
		\mathscr{A} &=&  N_K^2e^{-M_K(t_f-t_i)} \sum_{n}\frac{\langle\overline{K}^0|H_W|n\rangle\langle n|H_W|K^0\rangle}{M_K-E_n} 
\nonumber \\
&&\;\cdot \left( -T - \frac{1}{M_K-E_n} + \frac{e^{(M_K-E_n)T}}{M_K-E_n}\right).
\label{eq:integration_result}
\end{eqnarray}
Here $T=t_b-t_a+1$ is the time extent in lattice units of the integration volume and $N_K$ a known normalization factor associated with the interpolating operator $\overline{K}^0$. The differences $t_a-t_i$ and $t_f-t_b$ are assumed to be sufficiently large that only physical $\overline{K}^0$ and $K^0$ states appear in the initial and final states.  The coefficient of the term proportional to $T$ in Eq.~\eqref{eq:integration_result} provides a result for $\Delta M_K$:
\begin{equation}
\Delta M_K =  2 \sum_{n} \frac{\langle\overline{K}^0|H_W|n\rangle\langle n|H_W|K^0\rangle}{M_K-E_n}.
\label{eq:massdiff}
\end{equation}
The exponential terms coming from states $|n\rangle$ with $E_n>M_K$ in Eq.~\eqref{eq:integration_result} are exponentially decreasing as $T$ increases. These terms are negligible for sufficiently large $T$. For our small spatial volume and heavier than physical pion mass, there will be exponentially increasing terms coming from only $\pi^0$ and vacuum intermediate states. We evaluate the matrix element $\langle \pi^0 | H_W | K^0 \rangle$ and subtract this $\pi^0$ exponentially increasing term explicitly from Eq.~\eqref{eq:integration_result}. We also perform a subtraction for the $\eta$ state where the exponential decrease with increasing $T$ may be insufficient for it to be neglected.  This has a less than 10\% effect on the final result. For the vacuum state, we add a pseudo-scalar density, $\overline{s}\gamma^5 d$, to the weak Hamiltonian to eliminate the matrix element $\langle 0 | H_W + c_s \overline{s} \gamma^5 d| K^0 \rangle$. Since this pseudo-scalar density can be written as the divergence of an axial current, the final, physical mass difference will not be changed by adding this term. After the removal of these exponentially increasing terms, a linear fit at sufficiently large $T$ will determine $\Delta M_K$.

The $\Delta S=1$ effective Hamiltonian used in this calculation is
\begin{equation}
H_W=\frac{G_F}{\sqrt{2}}\sum_{q,q^{\prime}=u,c}V_{qd}V^{*}_{q^{\prime}s}(C_1Q_1^{qq^{\prime}}+C_2Q_2^{qq^{\prime}})
\label{eq:H_W}
\end{equation}
where $V_{qd}$ and $V_{q's}$ are Cabibbo-Kobayashi-Maskawa (CKM) matrix elements while
$\{Q_i^{qq{\prime}}\}_{i=1,2}$ are current-current operators, defined as:
\begin{equation}
\begin{split}
Q_1^{qq{\prime}}&=\bar{s}_i\gamma^\mu(1-\gamma^5)d_i \bar{q}_j\gamma^\mu(1-\gamma^5)q^{\prime}_j\\
Q_2^{qq{\prime}}&=\bar{s}_i\gamma^\mu(1-\gamma^5)d_j \bar{q}_j\gamma^\mu(1-\gamma^5)q^{\prime}_i\,.
\end{split}
\label{eq:operator}
\end{equation}
Since the Wilson coefficients $C_1$ and $C_2$ are calculated from the standard model in the continuum, we must relate our lattice operators to corresponding operators normalized in a continuum scheme.  We do this non-perturbatively using the Rome-Southampton, regularization invariant (RI) renormalization scheme~\cite{Martinelli:1994ty}. At present $C_1$ and $C_2$ have been computed to NLO in the $\overline{\mathrm{MS}}$ scheme~\cite{Buchalla:1995vs}.   We use a perturbative calculation of Lehner and Sturm, extending to our four-flavor case the results given in Ref.~\cite{Lehner:2011fz}, to convert these $\overline{\mathrm{MS}}$ values for $C_1$ and $C_2$ into the RI scheme.

\begin{figure}[!htp]
\begin{tabular}{cc}
	\hline
\includegraphics[width=0.2\textwidth]{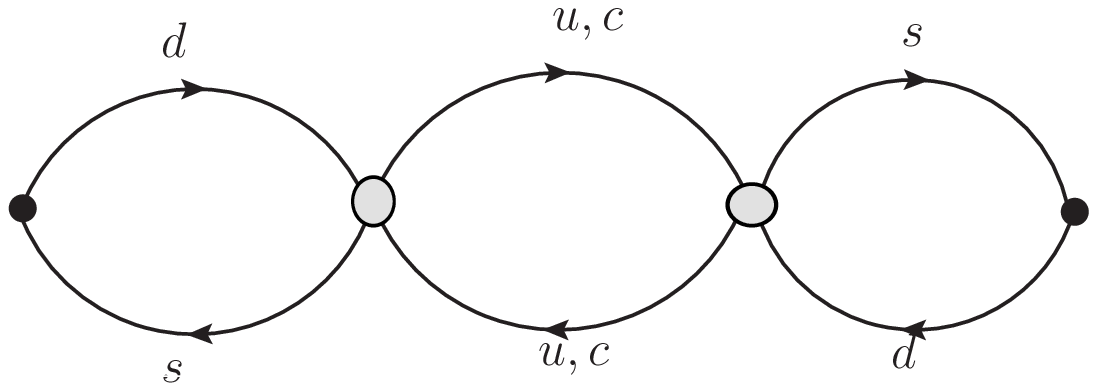} & \includegraphics[width=0.2\textwidth]{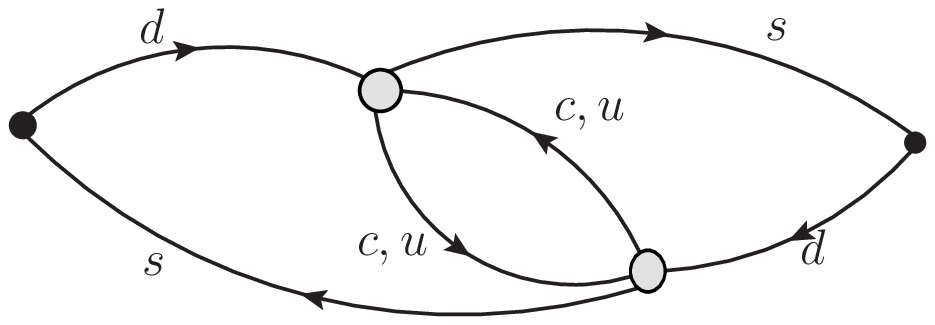} \\
Type 1 & Type 2 \\
\hline
\includegraphics[width=0.2\textwidth]{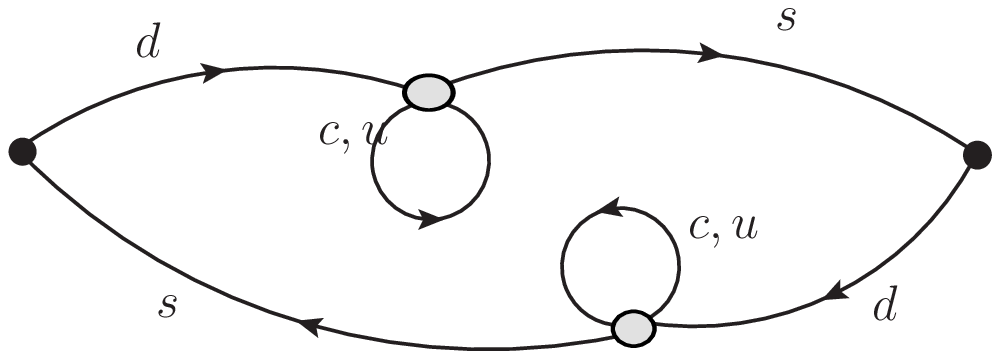} & \includegraphics[width=0.2\textwidth]{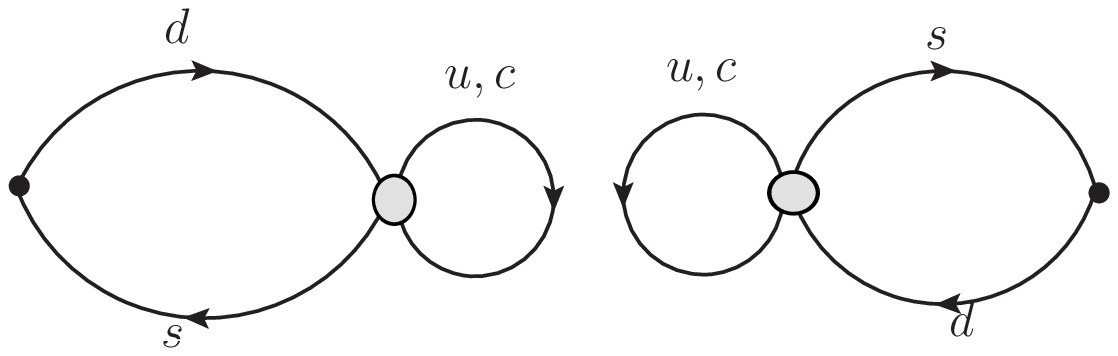}\\
Type 3 & Type 4\\
\hline
\end{tabular}
\caption{The four types of diagram contributing to the mass difference $\Delta M_K$. The shaded circles are the $\Delta S=1$ weak Hamiltonians. The black dots represent the kaon sources.}
\label{fig:diagrams}
\end{figure}

There are four types of diagrams, shown in Fig.~\ref{fig:diagrams}, that contribute to four-point correlator given in Eq.~\eqref{eq:integrated_correlator}. In our previous work~\cite{Christ:2012se}, we included only the first two types.  All diagrams are included in the present calculation. The disconnected, type 4 diagrams are expected to be the dominant source of statistical noise.

\section{Details of the calculation}
This calculation is performed on a lattice ensemble generated with the Iwasaki gauge action and 2+1 flavors of domain wall fermions~\cite{Allton:2008pn, Aoki:2010dy}. The space-time volume is $24^3\times64$ and the inverse lattice spacing $a^{-1}=1.729(28)$~GeV. The fifth-dimensional extent is $L_s=16$ and the residual mass is $m_{\mathrm{res}}=0.00308(4)$ in lattice units. The light and strange sea quark masses are $m_l=0.005$ and $m_s=0.04$, corresponding to pion and kaon masses $M_{\pi}=330$~MeV and $M_{K}=575$~MeV.   A valence charm quark with mass $m_c^{\overline{MS}}$(2~GeV) = 949~MeV provides GIM cancellation. We use 800 gauge configurations, separated by 10 time units.

We refer to Fig.~\ref{fig:int_correlator} to explain how this four point function is evaluated. We use Coulomb-gauge wall sources for the kaons. These two kaon sources are separated in time by $31$ lattice units. The two weak Hamiltonians are separated by at least $6$ time units from the kaon sources ($t_a-t_i$ and $t_f-t_b \ge 6$) so that the kaon interpolating operators will project onto physical kaon states. For type 1 and type 2 diagrams, we use the strategy of Ref.~\cite{Christ:2012se}: 64 propagators are computed using a point source on each of the 64 time slices.  The first of the two weak Hamiltonian densities is located at this point.  The propagators obtained with this point source are used to connect that Hamiltonian to the second Hamiltonian which can be summed over the full space-time region between $t_a$ and $t_b$.  For type 3 and type 4 diagrams, we use 64 random wall source propagators to construct the quark loops.  In order to reduce the noise coming from the random numbers, we use $6$ sets of random sources for each time slice, color and spin.  Thus, 4608 such random source propagators are computed for each gauge field configuration.  All the diagrams are averaged over all 64 time translations. For the light quark propagators, we calculate the lowest 300 eigenvectors of the Dirac operator and use low mode deflation to accelerate the light quark inversions.

\section{Results}
\begin{figure}[!htp]
\includegraphics[width=0.4\textwidth]{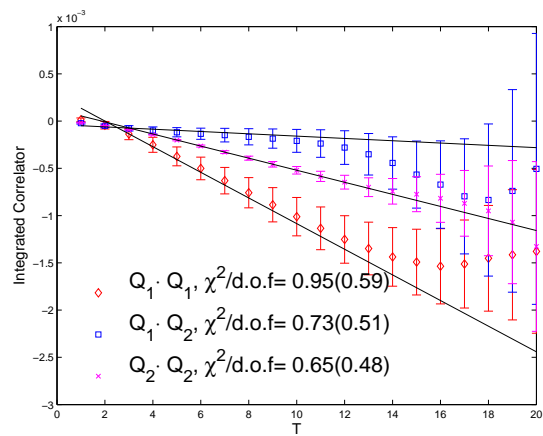} 
\caption{Integrated correlators for the three products of operators $Q_1\cdot Q_1$, $Q_1\cdot Q_2$ and $Q_2\cdot Q_2$. The three lines give the linear fits to the data in the time interval $[7, 20]$.}
\label{fig:int_corr}
\end{figure}

The results for the integrated correlators are shown in Fig.~\ref{fig:int_corr}. The three curves correspond to the three different operator combinations: $Q_1\cdot Q_1$, $Q_1\cdot Q_2$ and $Q_2\cdot Q_2$. The numbers are bare lattice results without any Wilson coefficients or renormalization factors. All the exponentially increasing terms have been removed from the correlators, so we expect a linear behavior for sufficiently large $T$. When $T$ becomes too large, the errors increase dramatically as should be expected since the disconnected diagrams have an exponentially decreasing signal-to-noise ratio. The straight lines correspond to linear fits to the data points in the range $[7,20]$. The $\chi^2/$d.o.f given in the figure suggest that these fits describe the data well.

\begin{figure}[!htp]
\includegraphics[width=0.4\textwidth]{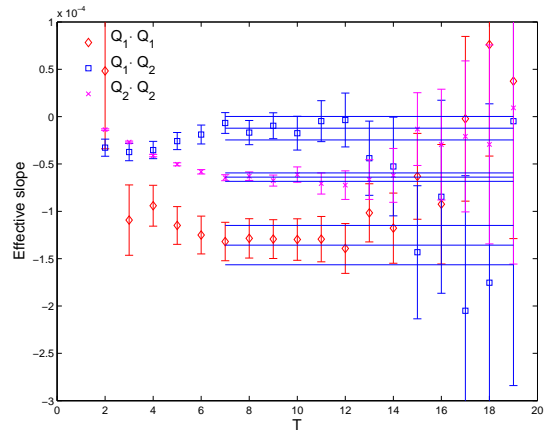} 
\caption{Effective slope plots for the three products of operators $Q_1\cdot Q_1$, $Q_1\cdot Q_2$ and $Q_2\cdot Q_2$.}
\label{fig:effslope_v2}
\end{figure}

Another method to check the quality of these fits is to plot the effective slope, in analogy to the effective mass plots used when determining a mass from a correlation function. The effective slope at a given time $T$ is calculated using a correlated linear fit to three data points at $T-1$, $T$ and $T+1$. In Fig.~\ref{fig:effslope_v2} we plot the effective slopes for the three different operator combinations. The horizontal lines with error bands give our final fitting results. For each operator combination we get good plateaus starting from $T=7$.

We have also tried different fitting ranges to see if our results depend sensitively on these choices.  We varied two parameters: the lower limit on the linear fitting range $T_{\mathrm{min}}$ and the minimum separation between the kaon sources and weak Hamiltonians $\Delta_{\mathrm{min}}$. We first fixed $\Delta_K=6$ and varied $T_{\mathrm{min}}$ from 7 to 9. The results are given in Table.~\ref{tab:fit_tmin}. While the central value of the fitting results is quite stable, the errors are sensitive to the choice of $T_{\mathrm{min}}$, which is caused by the disconnected diagrams. In Table.~\ref{tab:fit_deltak}, we give the results for fixed $T_{\mathrm{min}}=7$ but $\Delta_K$ varying from 6 to 8. Both the central values and the errors are very stable, suggesting that a separation of 6 is large enough to suppress excited kaon states.

\begin{table}[!htp]
	\caption{Results for the mass difference from each of the three operator products for different choices of $T_{\mathrm{min}}$ but with $\Delta_K$ fixed at 6.  All the masses are in units of $10^{-12}$~MeV.}
	\label{tab:fit_tmin}
	\begin{ruledtabular}
		\begin{tabular}{cccccc}
			$\Delta_K$ & $T_{\mathrm{min}}$ & $Q_1\cdot Q_1$ & $Q_1 \cdot Q_2$ & $Q_2 \cdot Q_2$ & $\Delta M_K$ \\
			\hline
			\multirow{3}{*}{6}	&	7	& 0.68(10) & -0.18(18) & 2.69(19) & 3.19(41)\\
			& 8  & 0.68(10) & -0.11(20) & 2.85(24) & 3.42(48) \\
		 &	9	& 0.68(11) & -0.18(25) & 2.69(34) & 3.18(63) \\
 \end{tabular}
 \end{ruledtabular}
\end{table}

\begin{table}[!htp]
	\caption{Fitting results for the mass difference from each of the three operator products for different choices of $\Delta_K$  but with $T_{\mathrm{min}}=7$. All the masses are in units of $10^{-12}$~MeV.}
	\label{tab:fit_deltak}
	\begin{ruledtabular}
		\begin{tabular}{cccccc}
			$T_{\mathrm{min}}$ & $\Delta_K$ & $Q_1\cdot Q_1$ & $Q_1 \cdot Q_2$ & $Q_2 \cdot Q_2$ & $\Delta M_K$ \\
			\hline
			\multirow{3}{*}{7} & 6 & 0.68(10) & -0.18(18) & 2.69(19) & 3.19(41) \\
		  &	7	& 0.68(10) & -0.20(18) & 2.64(19) & 3.13(41)\\
			 &	8	& 0.67(10) & -0.19(18) & 2.61(19) & 3.09(41) \\
 \end{tabular}
 \end{ruledtabular}
\end{table}

To check the calculation and refine our strategy for treating the exponentially growing single pion and vacuum contributions, we have varied the coefficient of the $\overline{s}\gamma^5 d$ term described above and introduced the similar $\overline{s} d$ operator .  Each operator is a total divergence and when added to $H_W$ should not change $\Delta M_K$.  In fact, $\Delta M_K$ did not change within errors as the coefficient of $\overline{s} d$ was varied.  We omit this term, since this gives the smallest statistical error.  In contrast, $\Delta M_K$ is very sensitive to $\overline{s}\gamma^5 d$.  If this term is omitted, the resulting exponentially growing vacuum contribution is two orders of magnitude larger than the previous linear term --- too large to be accurately subtracted.  Thus, we must use the $\overline{s}\gamma^5 d$ term to remove the vacuum intermediate state at the beginning.

In our previous work~\cite{Christ:2012se}, only the first two types of diagrams were included in the calculation. We can now determine the importance of these terms in our complete result. The contributions of the type 1 and 2 diagrams have small statistical errors and the coefficient of $T$ can be accurately determined from a linear fit using $T_{\mathrm{min}}=12$.  In Tab.~\ref{tab:type12}, we give the contribution to the three operator products from the type 1 and type 2 diagrams alone as well as the complete result. $\Delta M_K$ decreases by approximately a factor of two when the complete result is obtained, showing that there is large cancellation between the type 1 and 2  and the type 3 and 4 diagrams.  Since the type 3, ``double penguin'', graphs contribute less than 10\% to the final result, we find an unusually large contribution from the disconnected, type 4 diagrams. This is a surprisingly large failure of the``OZI  suppression''~\cite{Zweig:1964jf,*Okubo:1963fa,*Iizuka:1966fk}, naively expected for these disconnected diagrams. 

\begin{table}[!htp]
	\caption{Comparison of mass difference from type 1 and 2 diagrams only with that from all diagrams. All the numbers here are in units of $10^{-12}$~MeV.}
	\label{tab:type12}
	\begin{ruledtabular}
		\begin{tabular}{ccccc}
			 Diagrams & $Q_1\cdot Q_1$ & $Q_1 \cdot Q_2$ & $Q_2 \cdot Q_2$ & $\Delta M_K$ \\
			\hline
	Type 1,2		& 1.479(79) &  1.567(36) &  3.677(52) &  6.723(90) \\
			All  & 0.68(10) & -0.18(18) & 2.69(19) & 3.19(41) \\
 \end{tabular}
 \end{ruledtabular}
\end{table}

\section{Conclusions and outlook}
We have carried out the first, complete lattice QCD calculation of $\Delta M_K$.  However, our result is for a case of unphysical kinematics with pion, kaon and charmed quark masses of 330, 575 and 949~MeV respectively, each quite different from their physical values of 135, 495 and 1100~MeV. Our results is:
\begin{equation}
	\Delta M_K = 3.19(41)(96) \times 10^{-12} \quad\text{MeV}.
\end{equation}
Here the first error is statistical and the second an estimate of largest systematic error, the discretization error which results from including a 949~MeV charm quark in a calculation using an inverse lattice spacing of $1/a=1.73$~GeV . This 30\% estimate for the discretization error can be obtained either by simple power counting, $(m_c a)^2 = 0.30$, or from the failure of the calculated energy of the $\eta_c$ meson to satisfy the relativistic dispersion relation.  We find $(E_{\eta_c}^2(p)-p^2)/p^2 = 0.740(3)$ instead of 1.0 when evaluated at $p=2\pi/L$.

Our result for $\Delta M_K$ agrees well with the experimental value of $3.483(6)\times 10^{-12}$~MeV. However, since we are not using physical kinematics, this agreement could easily be fortuitous.  We emphasize that the objective of this first complete calculation is not a physical, standard model result for $\Delta M_K$ that should be compared with experiment but instead a demonstration that such a complete calculation is possible with controlled statistical errors.

To perform a calculation with physical kinematics and controlled systematic errors, two difficulties must be overcome. First, we need to perform the calculation on a four-flavor lattice ensemble with two or more, smaller lattice spacings. This would remove the difficult-to-estimate error associated with quenching the charm quark and allow the $O(m_c^2 a^2)$ discretization errors to be removed.  Second, we must perform a finite volume correction associated with $\pi-\pi$ re-scattering which will be needed for physical kinematics, when the two-pion threshold lies below the kaon mass. In this case, $\Delta M_K$ in infinite volume contains the principal part of the integral over the two-pion relative momentum, which can be substantially different from a finite-volume momentum sum.  A generalization of the Lellouch-Luscher method has been devised to correct this potentially large finite volume effect~\cite{Christ:2010zz} and a more general method has been presented in Ref.~\cite{Christ:2014qaa}.  Note, in future physical calculations with $L \approx 6$ fm there will be only one such two-pion state with energy well below $M_K$, contributing to $\Delta M_K$ on the few percent level.

Similar techniques can be used to determine the long distance contribution to $\epsilon_K$. However, the calculation of $\epsilon_K$ involves two additional complexities described in Appendix A of Ref.~\cite{Christ:2012se}. First, we must introduce new QCD penguin operators representing top quark effects. Second, an overall, logarithmic divergence must be removed from the lattice calculation using non-perturbative methods.   In summary, a full calculation of $\Delta M_K$ and $\epsilon_K$, including their long distance contributions, should be accessible to lattice QCD with controlled systematic errors within a few years, substantially increasing the importance of these quantities in the search for new phenomena beyond the standard model.

\begin{acknowledgments}
We thank our RBC and UKQCD colleagues for many valuable suggestions and encouragement and Guido Martinelli for helpful discussions. These results were obtained using the RIKEN BNL Research Center BG/Q computers at the Brookhaven National Laboratory.  N.C. and J.Y. were supported in part by US DOE grant DE-FG02-92ER40699, C.T.S. by STFC Grant ST/G000557/1, T.I. and A. S. by U.S. DOE contract DE-AC02-98CH10886 and T.I. also by JSPS Grants  22540301 and 23105715.
\end{acknowledgments}

\bibliography{citation}

\end{document}